\documentclass[prd,onecolumn,notitlepage,showpacs,preprintnumbers,amsmath,amssymb,nofootinbib,APS,10pt,superscriptaddress]{revtex4-1}

\usepackage{dcolumn}
\usepackage{bm}
\usepackage[applemac]{inputenc}
\usepackage[spanish,catalan,english]{babel}
\usepackage{amsmath,amssymb,amsfonts,latexsym,cancel}
\usepackage{graphicx}
\usepackage{color}
\usepackage{soul}
\usepackage{ulem}
\usepackage{hyperref}
\usepackage{slashed}

\allowdisplaybreaks

\newcommand{\be}{\begin{equation}}
\newcommand{\ee}{\end{equation}}
\newcommand{\bea}{\begin{eqnarray}}
\newcommand{\eea}{\end{eqnarray}}

\begin{document}
\title{{\bf Renormalization, running couplings and decoupling for the Yukawa model in curved spacetime}}

\author{Antonio Ferreiro}\email{antonio.ferreiro@ific.uv.es}
\author{Sergi Nadal-Gisbert}\email{sergi.nadal@uv.es}

\author{Jose Navarro-Salas}\email{jnavarro@ific.uv.es}
\affiliation{Departamento de F\'isica Te\'orica and IFIC, Centro Mixto Universidad de Valencia-CSIC. Facultad de F\'isica, Universidad de Valencia, Burjassot-46100, Valencia, Spain.}

\begin{abstract}

The decoupling of heavy fields as required by the Appelquist-Carazzone theorem plays a fundamental role in the construction of any effective field theory. 
However, it is not a trivial task to implement a renormalization prescription that produces the expected decoupling of massive fields, and it is even more difficult in curved spacetime. Focused on this idea, we consider the renormalization 
of the one-loop effective action for the Yukawa interaction with a background scalar field in curved space. We compute the beta functions within a generalized DeWitt-Schwinger subtraction procedure and discuss the decoupling in the running of the coupling constants. For the case of a quantized scalar field, all  the beta function exhibit decoupling, including also the gravitational ones. For a quantized Dirac field, decoupling appears almost for all the beta functions. We obtain the 
anomalous result that the mass of the background scalar field does not decouple. \\

 \end{abstract}

\date{\today}
\maketitle





\section{Introduction} 


The regularization and renormalization techniques in quantum field theory in curved spacetime are well established \cite{birrell-davies, fulling, Waldbook, parker-toms, hu-verdaguer, buchbinder-shapiro}. We have selfconsistent, covariant, and pragmatic procedures to evaluate the  expectation  value of the stress-energy tensor in physically reasonable states. The one-loop gravitational effective action is also a fundamental ingredient in the theory of quantized fields in curved spacetime. The main framework for the evaluation of the one-loop contribution to the effective action dates from the pioneer work by Schwinger \cite{schwinger51}, and it was generalized to curved spacetime by DeWitt \cite{DeWittbook, dewitt75}.  The DeWitt-Schwinger technique can be further adapted to implement different regularizations and subtraction methods in the evaluation of the renormalized effective action. While point-splitting is a preferred method for the renormalization of the stress-energy tensor, dimensional regularization and minimal  subtraction (MS) has been usually privileged in the calculations of the effective action and the associated evaluation of the running coupling constants \cite{parker-toms}. \\

The DeWitt-Schwinger proper-time expansion encounters an infrared divergence for massless fields, which can be naturally bypassed by  introducing  an upper cutoff in the proper-time integral  \cite{dewitt75}. The problem  can also be fixed  by replacing $m^2$ by an arbitrary mass scale $\mu^2$ in the conventional short-distance logarithmic term $\log m^2\sigma/2 $. By changing the energy scale $\mu$ to  $\mu'$ one can obtain the effective running of dimensionless coupling constants. [An old version of this argument for  the effective action of electrodynamics was pushed forward by Weisskopf \cite{Weisskopf}]. One can also obtain similar results using dimensional regularization and MS. In this approach, an arbitrary $\mu$ parameter is introduced to compensate the fictitious extra dimensions. Demanding that the bare parameter do not depend on $\mu$ produces a running in the renormalized couplings. \\ 

Another important issue concerns the decoupling in the running of the gravitational constants \cite{Babic2002,Sola-Shapiro}. This is a very elusive problem and it is far from trivial to design methods and strategies in the renormalization subtractions to implement the expected physical decoupling of heavy matter fields, in agreement with the Appelquist-Carazzone theorem \cite{APtheorem}\footnote{There are some situations were decoupling is violated. This typically happens on theories with spontaneously broken gauge symmetries \cite{Collins,Feruglio}.}. The overall idea of  effective field theory as applied in many branches of physics is largely based on the intuition of decoupling.  A field of mass $m$ cannot influence the physics on scales much larger than $m^{-1}$. Furthermore, in gravitational physics the fulfillment of the Appelquist-Carazzone theorem is essential to get a proper physical interpretation in the cosmic infrared regime. This is specially important for the cosmological constant problem and for the running of the Newton's gravitational constant  \cite{Carroll, Martin, Sola, review}. A partial list of 
works dealing with this issue in the gravitational context are \cite{franchinoreview,franchino19,gorbar-shapiro2003,gorbar-shapiro2004, antipin, markkanen}. 

An alternative way to introduce an energy scale $\mu$ in the DeWitt-Schwinger framework has been recently proposed in \cite{FN20} for quantized scalar fields coupled to an electromagnetic field and gravity (see also the related approaches \cite{FN, moreno-sola, BDNN}).  The advantage of the novel method is 
that it naturally decouples heavy massive fields from the  running. 
In this paper, we further extend this renormalization procedure to fermions and by including interactions, specifically, a Yukawa interaction between the quantized 
field and a classical and prescribed scalar field. This model has been recently studied by using dimensional regularization and minimal  subtraction (MS) \cite{Toms18,barra-buchbinder19,buchbinder19} (it has also been  extended including gauge fields in \cite{toms18gauge, toms19})\footnote{The decoupling in the Yukawa interaction with scalars in curved space was first studied in \cite{gorbar-shapiro2004} in the context where fields gain their masses due to spontaneous symmetry breaking.}. However MS subtraction does not fulfill the Appelquist-Carazzone theorem since the beta functions do not decouple, even in flat spacetimes. Here we consider the simplest case of a scalar background coupled to a massive quantized scalar and spinor fields via the Yukawa interaction. We will focus on the running of the coupling constants to see whether our generalized DeWitt-Schwinger method produces decoupling.\\


The paper is divided in two parts. In section \ref{SectionScalars} we introduce the generalized method of DeWitt-Schwinger subtractions with an arbitrary $\mu$ parameter. We deal with a quantized scalar $\varphi $ coupled,  via a Yukawa interaction $\frac{h^2}{2}\phi^2\varphi^2$, with a classical background scalar $\phi$. We compute the beta functions for the theory and observe decoupling for all 
coupling constants. In section \ref{SectionDirac} we analyze the generalized DeWitt-Schwinger adiabatic subtractions for the Dirac field. We consider a quantized Dirac field coupled to a scalar background via Yukawa interaction  $g_Y \bar\psi\psi \phi$. We compute the beta functions for the theory and discuss how decoupling appears for all the coupling constants except for the scalar mass parameter. Finally, we summarize our main conclusions in section \ref{conclusions}. We use units for which $c=1=\hbar$. Our sign conventions for the signature of the metric and the curvature tensor follow Ref. \cite{birrell-davies, parker-toms}. \\

  \section{Interaction with a quantized scalar field}\label{SectionScalars}
Consider a quantized real scalar field $\varphi$ coupled to a real scalar background $ \phi$ via the Yukawa interaction $\frac{h^2}{2}\phi^2\varphi^2$

\be\label{ClassicalAction}
S = \int d^4x \sqrt{-g}\left\{ -\Lambda + \frac{R}{16\pi G} + \frac12 \nabla^{\mu}\varphi\nabla_{\mu}\varphi - \frac12 \left(m^2 +\xi R \right) \varphi^2 - \frac{h^2}{2}\phi^2\varphi^2  + \frac12 \nabla^{\mu}\phi\nabla_{\mu}\phi - V(\phi)\right\} \ ,
\ee
where $m^2$ is the mass parameter for the quantized scalar, $\xi$ is the coupling of $\varphi^2$ to the Ricci scalar, and $V(\phi)$ is a general potential that can contain interactions between the background field and the curvature but is independent of the quantized scalar field $\varphi$. In order to visualize the divergences of the one-loop effective action  we can make use of the Feynman propagator $G_F$ satisfying
\be\label{ClassicalLagrangian1} \left(\Box_x+m^2+\xi R+h^2\phi^2\right)G_{\rm F}(x,x')=-|g(x)|^{-1/2}\delta(x-x').\ee
The effective action can be generated from this propagator  $S_{\rm eff}=-i \frac12 \operatorname{Tr} \log{(-G_{\rm F})}$. One can express the Feynman propagator as an integral in the proper time $s$ 
\be \label{GFs} G_{\rm F}(x, x') = -i \int_0^\infty ds \ e^{ -im^2 s} \langle x, s |  x', 0\rangle \ , \ee 
where $m^2$ is understood to have an infinitesimal negative imaginary part ($m^2\equiv m^2 -i\epsilon$). 
The heat kernel $\langle x, s |  x', 0\rangle$
can be expanded in powers of the proper time as follows 
 {\be \label{hks}\langle x, s |  x', 0\rangle =  i\frac{\Delta^{1/2} (x, x')}{(4\pi)^2(is)^2}  \exp {\frac{\sigma(x, x')}{2is}} \sum_{j=0}^\infty a_j (x, x')(is)^j \ , \ee 
where $\Delta(x, x')$ is the Van Vleck-Morette determinant and $\sigma(x, x')$ is  the proper distance along  the geodesic from $x'$ to $x$. Therefore, the effective Lagrangian, defined as $S_{\rm eff}=\int d^4x \sqrt{-g}L_{\rm eff}$, has the following  asymptotic expansion 
\be
L_{\rm eff}=\frac{i}{2(4\pi)^{2}}\sum^{\infty}_{j=0}a_j(x)\int^{\infty}_0e^{-is m^2}(is)^{j-3}ds \label{Leff}\ .
\ee
The first coefficients $a_n(x, x')$ are given, in the coincidence limit $x \to x'$, by  \cite{parker-toms}
\bea
a_0(x)=&&1 \ ,  \ \ \ \  \ \ a_1(x)=\frac16R-Q \nonumber\\
a_2(x)=&&\frac{1}{180}R_{\alpha\beta\gamma\delta}R^{\alpha\beta\gamma\delta}-\frac{1}{180}R^{\alpha\beta}R_{\alpha\beta} -\frac{1}{30}\Box R+\frac{1}{72}\ R^2 +\frac12 Q^2-\frac16 R Q +\frac16 \Box Q + \frac{1}{12} W_{\mu\nu}W^{\mu\nu}\ , \label{coef} 
 \eea
where for the case of a scalar field
we have $W_{\mu\nu}\equiv \left[\nabla_{\mu}, \nabla_{\nu} \right] = 0$ and $Q=\xi R+h^2 \phi^2$. The ultraviolet divergences of \eqref{Leff} are isolated in the first three terms of the DeWitt-Schwinger expansion. The renormalization procedure is set by directly subtracting the first three divergent terms appearing in the DeWitt-Schwinger expansion. \\

As already stressed in the introduction, the massless case inherits an infrared divergence. It can be avoided, for instance, by introducing an upper mass scale cutoff in the proper-time integral. Here we will follow an alternative strategy. We introduce a mass scale $\mu^2$ in the exponential term of the DeWitt-Schwinger expansion by writing $\sum_j a_j(x) e^{-ism^2} \to \sum_j \bar a_j(x) e^{-is(m^2+ \mu^2)}$ in (\ref{Leff}). This introduction of $\mu$ follows naturally once it is realized that for massless fields it is still mandatory to keep the exponential form. The DeWitt coefficients $a_j$ have to be  redefined by consistency.  As pointed in \cite{FN20}, this construction of the substraction terms will give rise to a decoupling of massive fields in the low energy limit for the case where the classical scalar field is absent, whereas the conventional drawback of dimensional regularization and MS is the absence of decoupling of heavy massive fields. We will check in the following if this is still valid in the Yukawa theory. \\

The DeWitt-Schwinger subtraction terms, upgraded with the introduction of the mass scale $\mu^2$, read as follows

\be
L_{\rm div}(\mu)=\frac{i}{2(4\pi)^{2}} \sum^{2}_{j=0}\bar{a}_j(x)\int^{\infty}_0e^{-is(m^2 + \mu^2)}(is)^{j-3}ds \ , \label{LdifmuScalars}
\ee
where $\bar{a}_0(x)=1, \bar{a}_1(x)=a_1(x)+\mu^2, \bar{a}_2(x)=a_2(x)+ a_1(x)\mu^2+\frac12\mu^4$ are fixed to keep consistency with each adiabatic order.  

\subsection{Running of the coupling constants and decoupling}
In order to obtain the beta functions from the DeWitt-Schwinger adiabatic subtractions we require that the total effective Lagrangian has to be independent of the arbitrary $\mu$ parameter. This requirement forces the parameters of the background Lagrangian $L_{\rm B}$ to run with the arbitrary $\mu$ scale. $L_{\rm B}$  is given by
\be\label{ClassicalLagrangian1}
L_{\rm B}= L_{\rm grav} +\frac{1}{2}Z \nabla^{\mu}\phi\nabla_{\mu}\phi - \frac{M^2}{2} Z \phi^2 -\frac{\xi_{\phi}}{2} R Z\phi^2  - \frac{\lambda}{4!}Z^2\phi^4  + \gamma_1 \Box Z\phi^2 \ ,
\ee
where
\be\label{GravitationalLagrangian}
L_{\rm grav} = -\Lambda +\frac{1}{2} \kappa R + \alpha_1R^2 
+ \alpha_2 R_{\mu\nu}R^{\mu\nu} + \alpha_3 R_{\mu\nu\alpha\beta}R^{\mu\nu\alpha\beta}+ \alpha_4  \Box R \ .
\ee
Here we have included only the necessary terms required for the theory to be renormalizable. These are given by the divergent terms of the heat kernel expansion. The coupling $Z$ will not receive any contribution from the scalar quantum fluctuations, therefore we can canonically normalize it to 1. The remaining couplings $\lambda(\mu), \kappa(\mu), \alpha_i(\mu), M(\mu)$, etc will inherit a dependence  on the mass scale $\mu$.\\

Because we are only interested in the $\mu$ dependent part of  \eqref{LdifmuScalars} and not in the full divergent term, we can make use of the difference between two values of $\mu$, i.e., 
\bea
L_{\rm div}(\mu)-L_{\rm div}(\mu_0)=&&\delta_{\Lambda}+\delta_{G}R+\delta_Q Q +\delta_{\sigma} a_2 
\ , \label{mum}
\eea
where
\bea \label{deltatemrs}
\delta_{\Lambda}&&=-\frac{1}{128\pi^2}\left\{-2m^2(\mu^2-\mu_0^2)+(\mu^4-\mu_0^4)+2m^4\log{\left(\frac{m^2 +\mu^2}{m^2+\mu_0^2}\right)}\right\} \nonumber \\
\delta_G&&= -\frac{1}{192\pi^2}\left\{-\mu_0^2+\mu^2-m^2\log{\left(\frac{m^2 +\mu^2}{m^2+\mu_0^2}\right)}
\right\} \nonumber \\
\delta_{\sigma}&&=-\frac{1}{32\pi^2}\log{\left(\frac{m^2 +\mu^2}{m^2+\mu_0^2}\right)}  \nonumber \\
\delta_Q&&=\frac{1}{32\pi^2}\left(\mu^2-\mu_0^2-m^2\log{\left(\frac{m^2 +\mu^2}{m^2+\mu_0^2}\right)} \right) \ .
\eea
The difference in the above expressions  with respect to dimensional regularization with MS is the appearance of  new terms  with powers of $\mu^4$ and $\mu^2$, which signal the presence of quartic and quadratic divergences. 
In MS, only the logarithmic divergences are reflected in the running of the couplings \cite{qftbook1,qftbook2,qftbook3}}.\\

If we demand that the physical one-loop renormalized Lagrangian $L_{\rm phys} = L_{\rm B}(\mu)+ L_{\rm eff} - L_{\rm div} (\mu)$ has to be $\mu$-independent, these leads to the relation
\be\label{RunningsEq}
L_{\rm B}(\mu) - L_{\rm B}(\mu_0) = L_{\rm div}(\mu)-L_{\rm div}(\mu_0) \ ,
\ee
which gives us a running for the background parameters. We can differentiate both sides of this equation and use the definition $\beta_\alpha \equiv \mu \frac{\partial \alpha}{\partial \mu} $ to obtain the beta functions for the couplings of \eqref{ClassicalLagrangian1} and \eqref{GravitationalLagrangian}:

\begin{alignat}{2}\label{betaDS1ScalarsDimenLess}
\beta_{\xi_{\phi}}&=\frac{h^2\bar\xi}{8\pi^2}\frac{\mu^2}{m^2+\mu^2} &\qquad \beta_{\alpha_1} & =-\frac{\bar\xi^2}{32\pi^2}\frac{\mu^2}{m^2+\mu^2} \nonumber \\
\beta_{\alpha_4} &=-\frac{\xi-\frac15}{96\pi^2}\frac{\mu^2}{m^2+\mu^2} &\qquad
\beta_{\alpha_2} & =\frac{1}{2880\pi^2}\frac{\mu^2}{m^2+\mu^2}\nonumber\\
\beta_{\alpha_3} & =-\frac{1}{2880\pi^2}\frac{\mu^2}{m^2+\mu^2} &\qquad
\beta{\gamma_1} & = -\frac{h^2}{96\pi^2}\frac{\mu^2}{m^2+\mu^2} \nonumber \\
\beta_{\lambda} & =\frac{3 h^4}{4\pi^2}\frac{\mu^2}{m^2+\mu^2} &\qquad
\beta_{\Lambda} & =\frac{1}{32\pi^2}\frac{\mu^6}{m^2+\mu^2} \nonumber \\
\beta_{\kappa} & =\frac{\bar\xi}{8\pi^2}\frac{\mu^4}{m^2+\mu^2}& \qquad
\beta_{M^2}& =-\frac{h^2}{8\pi^2}\frac{\mu^4}{m^2+\mu^2}  \ ,
\end{alignat}
where we have defined $\bar\xi=\left(\xi-\frac16\right)$. The runnings for the  gravitational couplings  are indeed the same as in for the free field theory \cite{FN20}, moreover we get decoupling for the rest of beta functions linked to the background scalar couplings. It is easy to check from these results that all coupling constants do indeed decouple in the infrared regime, i.e., $m^2 \gg \mu^2$, even for dimensionfull couplings.  We stress here that it is far from trivial to obtain decoupling for all coupling constants \cite{franchinoreview,franchino19,gorbar-shapiro2003,antipin}. We think it is quite remarkable that the above procedure has been able to achieve complete decoupling for all couplings of the theory. Notice that the factors of the form $\mu^2/(m^2 + \mu^2)$ also arise in the hierarchy of beta functions in the Wilsonian renormalization approach for a scalar field theory \cite{Hollowood}. 
 It is also important that the  generalized DeWitt-Schwinger subtraction method gives both the physical ultraviolet and infrared behaviour for the dimensionless couplings. One can check that our result is consistent with dimensional regularization with MS for the expected high energy behavior ($\mu^2 \gg m^2$) of the dimensionless constants \cite{Toms18,barra-buchbinder19}.\\ 

The running of the dimensionfull couplings, the cosmological $\Lambda$ and Newton's gravitational constants $G$ and the scalar background mass $M^2$ are given by ($\Lambda= \Lambda_c/8\pi G$, where $\Lambda_c$ is the traditional cosmological constant).

\bea \label{rlambda}
\Lambda(\mu)&=&\Lambda_0-\frac{1}{128\pi^2}\left(-(\mu^4-\mu_0^4)+2m^2(\mu^2-\mu_0^2) - 2m^4\log{\left(\frac{m^2+\mu^2}{m^2+\mu_0^2}\right)}\right)\ , \label{ccmuScalars}  \\
G(\mu)&=&\frac{G_0}{1+\bar\xi\frac{ G_0}{2\pi}\left(\mu^2-\mu_0^2-m^2\log{\left(\frac{m^2+ \mu^2}{m^2+\mu_0^2}\right)}\right)} \ , \label{rG} \\
M^2(\mu)&=&M_0^2-\frac{h^2}{16\pi^2}\left((\mu^2-\mu_0^2)-m^2\log{\left(\frac{m^2+ \mu^2}{m^2+\mu_0^2}\right)}\right)\ . \label{rM}
\eea
The new $\mu^2$ and $\mu^4$ terms which signals the presence of quadratic and quartic divergences are the responsible for the decoupling of the dimensional constants. Expanding these expressions for large $m^2$, we see how the couplings exactly decreases proportional to $\frac{1}{m^2}$ ($\Lambda(\mu)  \propto \frac{\mu^6}{m^2}$, $G(\mu)\propto \frac{\mu^4}{m^2}$ and $M^2(\mu)  \propto \frac{\mu^4}{m^2}$). Therefore, we have obtained a physically consistent renormalization flow which agrees with decoupling for $m^2 \gg \mu^2$ for all the parameters, even the dimensionfull ones.  

\section{renormalization for Dirac fields}\label{SectionDirac}

In this section we extend the computation of the renormalization method to take into account a quantized Dirac field coupled via Yukawa interaction with a classical scalar background. 
Let us assume a quantized Dirac field in curved spacetime with a Yukawa coupling with the action given by
\be\label{YukawaFermionsClassicalAction}
S=\int d^4x \sqrt{-g}\left( -\Lambda + \frac{R}{16\pi G} + \bar\psi \left( i\gamma^{\mu}\nabla_{\mu} - m\right)\psi - g_Y\phi\bar\psi\psi   + \frac12 \nabla^{\mu}\phi\nabla_{\mu}\phi - V(\phi) \right) 
\ , \ee
where $\phi$ is treated as a classical background. The covariant derivative $\nabla_\mu$ acting on the Dirac field  is defined as the  ordinary derivative plus the spin connection term. $\gamma^\mu(x)$ are the curved space Dirac matrices $\gamma^\mu(x)= e^{\ \mu}_a \gamma^a$, defined in terms of the usual Dirac matrices in Minkowski space $\gamma^a$ and the vierbein $e^{\ \mu}_a$.  A previous analysis of this theory in the context of adiabatic regularization can be seen in \cite{FDNT}.\\

The effective action induced by the quantum fluctuations of the Dirac field is given by
\be  S_{\rm eff}=i \operatorname{Tr} \log (-S_F) , \ee
where the Feynman propagator for the fermionic field satisfies the equation
\be\label{DiracPropEq}
 \left[ i \gamma^{\mu}\nabla_{\mu}-m-g_{\rm Y}\phi\right]S_F(x,x')=-|g(x)|^{-1/2} \delta (x-x') . \ee
This is a linear order differential equation. However, in order to use the DeWitt-Schwinger expansion it is needed a Klein-Gordon second order type equation for the Green function. One can manipulate the above expression to rewrite the effective action in the following convenient form \cite{parker-toms} $S_{\rm eff}^{(1)}= \frac{1}{2}i \operatorname{Tr} \log (-G_F) $ 
 where $G_F$ is defined such that 
 \be S_F(x,x')= \left[ i \gamma^{\mu}\nabla_{\mu}+m+g_{\rm Y}\phi\right]G_F(x,x') \ .  \ee 
 The advantage of introducing the new Green function $G_F(x,x')$ is that it obeys the second-order differential  equation
   \be \label{KG1Ferm}(\Box_x  + m^2 + Q) G_{\rm F}(x, x') = -|g(x)|^{-1/2} \delta (x-x') \ , \ee
thus we can use the DeWitt-Schwinger expansion as for the scalar field. In this case $Q$ is given by
\be Q(x) = \frac14 R(x) + i g_Y \gamma^{\mu}\nabla_{\mu}\phi(x) + g_Y^2 \phi^2 (x)+ 2 g_Y m \phi(x)  \label{qfermion}\ . \ee
%
%
Notice the appearance of a term proportional to $m$ in the above expression for $Q$. This contrast with the obtained expression for $Q$ in the scalar case ($Q=\xi R+h^2 \phi^2$). Generically, all DeWitt coefficients  $a_n$ (or $\bar a_n$) are local geometrical quantities independent of the mass of the field. The Yukawa interaction for Dirac fermions introduces a mass-dependent term in the expression for $Q(x)$. This will have important consequences.\\

Following the same calculation as in the previous section, we get the expression for the subtraction terms with the first coefficients $a_0(x), a_1(x)$ and $ a_2(x)$ of the asymptotic expansion
\be
L_{\rm div}(\mu)=\frac{-i}{2(4\pi)^{2}}\sum^{2}_{j=0}\operatorname{tr}\bar{a}_j(x)\int^{\infty}_0e^{-is(m^2 + \mu^2)}(is)^{j-3}ds \ , \label{Ldifmu3}
\ee
where $\operatorname{tr}\bar{a}_j(x)$, takes the trace of the spinor indices acting on the coefficients $\bar{a}_j(x)$. We still have the modified DeWitt coefficients $\bar{a}_0(x)=1$, $\bar{a}_1(x)=a_1(x)+\mu^2$,  and $\bar{a}_2(x)=a_2(x)+\left(\frac{1}{6}R - Q\right)\mu^2+\frac12\mu^4$
related to the DeWitt coefficients  \eqref{coef} but now with $Q$ given by \eqref{qfermion}  and $W_{\mu\nu} = [\nabla_{\mu},\nabla_{\nu}]= -\frac{1}{8} R_{\mu\nu a b}\left[\gamma^a, \gamma^b \right] $.\\

\subsection{Running of the coupling constant and decoupling}
As for the scalar case, in order to obtain the beta functions from the DeWitt-Schwinger adiabatic subtractions method it is enough to study the divergent contribution to the effective action induced by the quantum fluctuations of the Dirac field. The required terms for renormalization that must contain the background Lagrangian are obtained from the divergent part of the effective action \eqref{Ldifmu3}
\be\label{ClassicalL}
L_{\rm B}= L_{\rm grav} +\frac{1}{2} Z \nabla^{\mu}\phi\nabla_{\mu}\phi - \frac{M^2}{2} Z \phi^2 
-\frac{\xi}{2} R Z\phi^2 -\tau Z^{1/2}\phi - \frac{\eta}{3!}Z^{3/2}\phi^3 - \frac{\lambda}{4!}Z^2\phi^4 - \xi_1R Z^{1/2}\phi + \gamma_1 \Box Z\phi^2 +  \gamma_2\Box Z^{1/2} \phi \ ,
\ee
where $L_{\rm grav}$ was given in \eqref{GravitationalLagrangian}. In the Dirac case, $Z$ gets a contribution from the quantum fluctuations of the Dirac field. Therefore, it has a running that can be related to a running of the field $\phi$ by a reparametrization. We write $Z = 1 + \delta Z $ as usual, to take into account canonical normalization and the one-loop correction. For simplicity, we introduce new primed couplings to absorb $Z$, except for the kinetic term, where we leave $Z$ explicitly.
\be\label{ClassicalLPrimed}
L_{\rm B}= L_{\rm grav} + \frac{1}{2} Z\nabla^{\mu}\phi\nabla_{\mu}\phi - \frac{M'^2}{2} \phi^2 - \frac{\xi'}{2} R \phi^2 -\tau'\phi - \frac{\eta'}{3!}\phi^3 - \frac{\lambda'}{4!}\phi^4 - \xi'_1R \phi + \gamma'_1 \Box \phi^2 + \gamma'_2\Box  \phi \ .
\ee
We evaluate the difference of the divergent contributions between  two arbitrary ($\mu, \ \mu_0$) renormalization points
\bea
L_{\rm div}(\mu)-L_{\rm div}(\mu_0)=&&\delta_{\Lambda}+\delta_{G}R+\delta_Q \operatorname{tr} Q +\delta_{\sigma} \operatorname{tr} a_2 
\ , \label{mum}
\eea
where we also get quartic $\mu^4$ and quadratic $\mu^2$ terms as in the scalar case

\bea \label{naiver}
\delta_{\Lambda}&&=\frac{1}{32\pi^2}\left(\mu^4-\mu_0^4+ 2m^2(\mu_0^2-\mu^2)+ 2m^4\log{\left(\frac{m^2+\mu^2}{m^2+\mu_0^2}\right)}\right) \ ,\nonumber \\
\delta_G&&=\frac{1}{48\pi^2}\left(-\mu_0^2+\mu^2- m^2\log{\left(\frac{m^2+\mu^2}{m^2+\mu_0^2}\right)}\right) \ ,
\nonumber \\
\delta_{\sigma}&&=\frac{1}{32\pi^2}\log{\left(\frac{m^2+\mu^2}{m^2+\mu_0^2}\right)} \ ,\nonumber \\
\delta_Q&&=\frac{1}{32\pi^2}\left(\mu_0^2-\mu^2 + m^2\log{\left(\frac{m^2+\mu^2}{m^2+\mu_0^2}\right)}\right) \ .
\eea 
As before, we impose that the total one-loop renormalized Lagrangian $L_{\rm phys}=L_{\rm B}(\mu)+ L_{\rm eff} - L_{\rm div} (\mu)$ must be independent of the value of $\mu$. This gives us the running for the primed couplings. By direct differentiation and keeping one-loop order $\mathcal{O}(\hbar)$, it is straightforward to obtain the beta functions of the unprimed original couplings 

\bea\label{BetaFunctStructureOne-loop}
\beta_{M^2} && = \beta_{M'^2}- M^2\beta_{Z}    \nonumber \\
\beta_{\xi} && = \beta_{\xi'}- \xi \beta_{Z}  \nonumber \\
\beta_{\xi_1} && = \beta_{\xi'_1}- \frac 12 \xi_1 \beta_{Z} \nonumber \\
\beta_{\tau} && = \beta_{\tau'}- \frac{\tau}{2}\beta_{Z}  \nonumber \\
\beta_{\eta} && = \beta_{\eta'}- \frac{3}{2}\eta \beta_{Z}  \nonumber \\
\beta_{\lambda} && = \beta_{\lambda'}- 2\lambda \beta_{Z} \nonumber \\
\beta_{\gamma_1} && = \beta_{\gamma'_1}- \gamma_1\beta_{Z}  \nonumber \\
\beta_{\gamma_2} && = \beta_{\gamma'_2}- \frac12 \gamma_2\beta_{Z} \ .\nonumber 
\eea
%
%
The result for all the beta functions are listed below in \eqref{betaFermionsAdimen} and \eqref{betaFermionsDimensional}. Let us analyze the infrared and ultraviolet regimes of these runnings with more detail. As a representative of the dimensionless couplings we take the scalar wavefunction $Z$ with the following beta function
\be
\beta_{Z}=-\frac{g_Y^2}{4\pi^2}\frac{\mu^2}{m^2+\mu^2} \label{betaq2}\ .
\ee
In the ultraviolet regime, $\mu \gg m$ we recover the result from dimensional regularization with MS \cite{Toms18, barra-buchbinder19}.
\be \beta_{Z}=-\frac{g_Y^2}{4\pi^2}\frac{\mu^2}{m^2+\mu^2} \to_{\mu \gg m}  -\frac{g_Y^2}{4\pi^2} \ ,\ee
while for the infrared regime, $\mu \ll m$, we find  decoupling
\be \beta_{Z}=-\frac{g_Y^2}{4\pi^2}\frac{\mu^2}{m^2+\mu^2} \to_{\mu \ll m}  -\frac{g_Y^2}{4\pi^2}\frac{\mu^2}{m^2} \ .\ee
Similarly, we also find decoupling when $m \gg \mu$ for all dimensionless coupling constants. Moreover, and as a consistent check of our method we see that for $\mu \gg m$ the dimensionless beta functions agree with the behavior obtained from dimensional regularization \cite{Toms18,barra-buchbinder19}. The dimensionless beta functions are:
\begin{alignat}{2}\label{betaFermionsAdimen}
 \beta_{\xi} & =-\frac{g_Y^2}{24\pi^2}\frac{\mu^2}{m^2+\mu^2}\left(1-6\xi\right) &\qquad \beta_{\lambda}&=-\frac{ g_Y^2 }{8\pi^2}\frac{\mu^2}{m^2+\mu^2}\left(24g_Y^2 - 4\lambda\right)    \nonumber\\
 \beta_{\alpha_1}& =\frac{1}{1152\pi^2}\frac{\mu^2}{m^2+\mu^2} &\qquad
\beta_{\alpha_2}&=-\frac{1}{720\pi^2}\frac{\mu^2}{m^2+\mu^2}  \nonumber \\
\beta_{\alpha_3}&=-\frac{7}{5760\pi^2}\frac{\mu^2}{m^2+\mu^2} &\qquad
\beta_{\alpha_4}&= \frac{1}{480\pi^2}\frac{\mu^2}{m^2+\mu^2}  \nonumber \\
\beta_{\gamma_1}& =\frac{g_Y^2}{4\pi^2}\frac{\mu^2}{m^2+\mu^2}\left(\frac16 +\gamma_1\right) &\qquad \beta_{\gamma_2}&=\frac{g_Y}{8\pi^2}\frac{\mu^2}{m^2+\mu^2} \left(\frac23 m + \gamma_2\right) \ .
\end{alignat}
Things are more involved for the dimensionfull constants:  
\begin{alignat}{2}\label{betaFermionsDimensional}
\beta_{\Lambda}&=-\frac{1}{8\pi^2}\frac{\mu^6}{m^2+\mu^2} &\qquad \beta_{\xi_1}&=\frac{g_Y}{8 \pi^2}\frac{\mu^2}{m^2+\mu^2}\left(-\frac{m}{3} + g_Y\xi_1\right) \nonumber \\
\beta_{\kappa}&=-\frac{1}{24\pi^2}\frac{\mu^4}{m^2+\mu^2} &\qquad \beta_{\tau}&=-\frac{g_Y  }{8\pi^2}\frac{\mu^2}{m^2+\mu^2}\left(4m\mu^2 - \tau g_Y \right) \nonumber \\
\beta_{M^2}&=\frac{g_Y^2}{8\pi^2}\frac{\mu^2}{m^2 + \mu^2}\left(+4\mu^2-8m^2 + 2M^2 \right) &\qquad \beta_{\eta}&=-\frac{ g_Y^3 }{8\pi^2}\frac{\mu^2}{m^2+\mu^2} \left(24 g_Y m - 3\eta \right) \ .
\end{alignat}
We can immediately see that decoupling appears when $m \gg \mu$ for all the dimensionfull parameters except for the scalar mass term $M^2$, where the beta function essentially reproduces the value obtained via dimensional regularization with MS times a factor of order $\mu^2/m^2$. The origin of the term proportional to $-8m^2$ can be retrieved from the term $2g_Y m \phi$ in $Q$.  The latter (disturbing) result shows that finding decoupling for all the coupling constants of a given theory is indeed a non-trivial task, as already emphasized in \cite{franchinoreview}.\\

Let us focus now on the finite expression for the running  of the Newton gravitational constant $G$, the cosmological constant $\Lambda$ and the scalar mass term $M'^2$. We get the following runnings for these parameters from the above analysis 
\bea \label{rlambda}
\Lambda(\mu)&=&\Lambda_0-\frac{1}{32\pi^2}\left((\mu^4-\mu_0^4)-2m^2(\mu^2-\mu_0^2) 
+ 2m^4\log{\left(\frac{m^2+\mu^2}{m^2+\mu_0^2}\right)}\right)\ , \label{ccmu}  \\
G(\mu)&=&\frac{G_0}{1-\frac{ G_0}{6\pi}\left(\mu^2-\mu_0^2-m^2\log{\left(\frac{m^2+ \mu^2}{m^2+\mu_0^2}\right)}\right)} \ , \label{rG} \\
M'^{2}(\mu)&=&M'^{2}_0+\frac{g^2_Y}{4\pi^2}\left((\mu^2-\mu_0^2)-3m^2\log{\left(\frac{m^2+ \mu^2}{m^2+\mu_0^2}\right)}\right)\ .\label{rM}
\eea
For the gravitational couplings, at large values of the scale $\mu \gg m$, the mass $m$ can be ignored, while heavy particles decouple when $m \gg \mu$ so there is no running in the infrared regime. However, the mass parameter $M'^2$ still gets a running dependence on $\mu^2$ when $m \gg \mu$ (the mass coupling $M^2 = M'^2/Z$ will exhibit the same anomalous behaviour). These features can be seen explicitly in the running of the parameters by expanding the above expressions for large $m^2 / \mu^2$
\bea \label{Frlambda}
\Lambda(\mu)&\sim &\Lambda_0 + \frac{\text{$\mu_0 $}^6-\mu ^6}{48 \pi ^2 m^2}+O\left(m^{-3}\right) \ ,\label{Fccmu}  \\
G(\mu)&\sim & G_0 + \frac{G_0\left( \mu ^4-\text{$\mu_0 $}^4\right)}{12 \pi  m^2}+O\left(m^{-3}\right) \ ,\label{rG} \\
M'^2(\mu) & \sim & M'^2_0 - \frac{g^2_Y \left(\text{$\mu_0 $}^2-\mu ^2\right)}{2 \pi ^2}+\frac{3 \left(g^2_Y \left(\text{$\mu_0$}^4-\mu ^4\right)\right)}{8 \pi ^2 m^2}+O\left(m^{-3}\right) \ . \label{FrM}
\eea

\section{Conclusions and final comments}  \label{conclusions} 

We have computed the one-loop beta functions for the Yukawa interaction in curved spacetime in a generalized DeWitt-Schwinger framework. The generalization involves the introduction of an  arbitrary but necessary mass scale $\mu$ parameter in the DeWitt-Schwinger adiabatic subtraction method. From this, we derive the renormalization group flow of the couplings.  We have shown in full detail how decoupling of heavy massive fields naturally appears, for both fermionic and scalar sector. The most important result is the appearance of non-logarithmic runnings for the dimensionfull couplings, like the cosmological constant and Newton's constant. These new quadratic and quartic dependence on $\mu$ are indeed responsible of decoupling. Finally, we would like to stress that although decoupling remarkably appears for almost all the coupling constants, the scalar mass coupling still gets a quantum contribution from the massive Dirac field in the low energy regime. The disturbing behaviour for the running of the external scalar mass  shows indeed the extreme difficulty of obtaining a consistent renormalization procedure that incorporates full decoupling. According to our analysis of the fermionic Yukawa model, the reason seems to be located in the massive linear term $2 g_Y m \phi $ in $Q$. We are currently investigating how to overcome this latter difficulty. \\


{\it Acknowledgments.--} 
 We thank I. L. Shapiro for useful comments. This work has been  partially supported  by Spanish Ministerio de  Economia,  Industria  y  Competitividad  Grants  No. FIS2017-84440-C2-1-P (MINECO/FEDER, EU) and No.  FIS2017-91161-EXP, and also by the project PROMETEO/2020/079 (Generalitat Valenciana).  A. F. is supported by the Severo Ochoa Ph.D. fellowship, Grant No. SEV-2014-0398-16-1, and the European Social Fund. S. N. is supported by the Universidad de Valencia, within the Atracci\'o de Talent Ph.D fellowship No. UV-INV- 506 PREDOC19F1-1005367.

\end{document}